
\magnification 1200
\def\cM {{\cal{M}}}
\def\eq  {\simeq}
\def\ggg{\vrule width0.5pt height5pt depth1pt}
\def\pp{{{ =\hskip-3.75pt{\ggg}}\hskip3.75pt }}
\def\cL{{\cal {L}}}

\font\mybb=msbm10 at 10pt
\def\bb#1{\hbox{\mybb#1}}
\def\RN {\bb{R}}
\def\ZN {\bb{Z}}


 \rightline{June 1994~~~}
\vskip 1cm

\centerline{\bf {GLOBAL ASPECTS OF SYMMETRIES IN SIGMA MODELS WITH TORSION}}
\vskip 3cm
\centerline{{\bf G. Papadopoulos}\footnote{$^*$}{Address
after 1st October: DAMTP,
University of Cambridge, England.}}
\vskip 0.5cm
\centerline{II. Institute for Theoretical Physics}
\centerline{University of Hamburg}
\centerline{Luruper Chaussee 149}
\centerline{22761, Hamburg}
\centerline{Germany}
\vskip 0.5cm
\centerline{and}
\vskip 0.5cm
\centerline{Blackett Laboratory}
\centerline{Imperial College}
\centerline{Prince Consort Road}
\centerline{London SW7 2BZ, UK}
\vskip 2.5cm

\centerline{ABSTRACT
\footnote{$^\dagger$} {To appear in the proceedings of
{\sl Geometry of Constrained Dynamical Systems} workshop, Newton
Institute,  Cambridge, June 1994.}}
\medskip

It is shown that non-trivial topological sectors can prevent the quantum
mechanical implementation of the symmetries of the classical
field equations of sigma models
with torsion.  The associated anomaly is computed,
and it is shown  that it depends on the
homotopy class of the topological sector of the
theory and the group action on the sigma
model manifold that generates the symmetries of the classical field equations.

\vfill\eject

\noindent {\bf 1. Introduction }
\medskip

Two-dimensional sigma models have been extensively
studied the last ten years because of
their applications to string theory and the
theory of integrable systems.  More recently,
attention has been focused on sigma models
with symmetries.  Such sigma models arise
naturally in the investigation of sigma model
duality (see for example refs. [1]) and the
study of supersymmetric sigma models with potentials [2,3,4].

The fields of a sigma model are maps $\phi$ from the
two-dimensional space-time $\Sigma$ into
a Riemannian manifold $(\cM, h)$ which is called
sigma model manifold or target space, and
the couplings are tensors on
$\cM$. In the following, we will focus on  sigma models with
couplings a metric $h$, a scalar function $V$, and an
antisymmetric two-form $b$. The form $b$ may be locally
defined on $\cM$ and the part of the
sigma model action $I$ that contains this coupling is
called Wess-Zumino (WZ) term [5] or WZ
action.   The action $I$ is either {\sl locally} or {\sl globally} defined
depending on whether the WZ term $b$ is a locally or
globally defined form correspondingly.
The classical field equations, however, are always {\sl globally} defined on
$\cM$ because the coupling $b$ enters in them through its exterior derivative
$H={3\over2}db$ and $H$ is a {\sl globally} defined
closed three-form on $\cM$ (see
section 3).  From the classical field  theory point of view,
the latter is enough to
guarantee that the above theory is covariant under
reparameterisations of the sigma model
manifold and so well defined.  In the path integral
quantisation of this theory, however,
$\exp( 2\pi i I)$ is required to be globally defined as well,
{\sl i.e.}\   the sigma model action $I$ must be globally
defined up to an integer.  If the
action $I$ is not globally defined, this
additional requirement leads to a certain
quantisation condition for the WZ coupling $b$ [6,7].

The symmetries of sigma models that we will
examine are those induced by vector fields on
the sigma model manifold.  A sufficient condition
for the transformations generated by
vector fields on the sigma model manifold to be
symmetries of the {\sl field equations} is to
leave the  tensors $h,V,H$ invariant.  In the quantum theory though, we
need to know the conditions for the invariance of the
sigma model {\sl action} under the
above transformations.  This is straightforward for the
terms in the action that contain
couplings $h$ and $V$; the conditions are the same as those for the invariance
of the associated terms in the field equations.
{\sl New} conditions are required though
for the invariance of the WZ action in addition to those necessary
for the invariance of the corresponding term
in the field equations; I will call these
new conditions anomalies.  This is due to the fact that  the WZ action may be
locally defined on
$\cM$ and  the symmetries of the field equations may leave this action
invariant up to surface terms that cannot
be integrated away. (The latter can happen even in
the special case where $b$ {\sl is} globally defined on $\cM$).

The main point of this talk is to present the
conditions under which the action of a sigma
model with a WZ term is invariant (up to an integer)
under the transformations  generated by
a group
$G$ acting on the sigma model manifold $\cM$.
I will then introduce the (1,1)-supersymmetric
two-dimensional massive sigma model and show that
the above anomaly cancels for the
symmetries that arise naturally in this model.

The conditions for the invariance of the WZ
action have been presented before by the
author in ref. [8]. However the original
publication does not contain the proof of key
statements regarding these conditions and there is no mention  of the
conditions
that are necessary for the WZ action to be
invariant under infinitesimal transformations.
These will be included here.  The action of massive
(1,1)-supersymmetric sigma models was given
in ref. [3] in collaboration with C.M. Hull
and P.K. Townsend.

In section two, I will briefly review the conditions for the symmetries of the
equations of motion of  a charged particle
coupled to a magnetic field to be implemented in
the quantum theory of the system.  In section
three, I will give the action of a bosonic
sigma model with a WZ term and a scalar potential,
and the conditions for its field equations
to be invariant under transformations generated
by a group acting on the sigma model
manifold.  In section four, I will present a global
definition of the WZ action and give the  conditions
for this action to be invariant under
the symmetries of its field equations, and in
section five, I will discuss the symmetries of
supersymmetric massive sigma models.

\bigskip
{\bf 2. A Quantum Mechanical Model}
\medskip

It is instructive to present the conditions
under which the symmetries of the equations of
motion of a charged particle coupled  to a
magnetic field can be implemented quantum
mechanically [9].  This is because
there is a close relation between these conditions and
some of the conditions that I will derive in section 4 for the case of the
two-dimensional sigma model with a WZ term. The action of a charged particle
coupled to a magnetic field $b$ is
$$
I=\int dt\ \ {1\over2} h_{ij}\partial_t \phi^i
\partial_t \phi^j+ b_i\partial_t \phi^i\ ,
\eqno (2.1)
$$
where $t$ is a parameter of the world-line,
$\phi$ are the co-ordinates of the particle and
$h$ is the metric of the manifold
$\cM$ in which the particle propagates. The
coupling $b$ is a locally defined one-form on
$\cM$ with patching conditions $b_1=b_2+da_{12}$
on the intersection $U_1\cap U_2$ of any
two open sets
$U_1,U_2$ of $\cM$, and $a_{12}$ is a
function on $U_1\cap U_2$.  Because of these patching
conditions the last term of the above action
is locally defined on $\cM$.  The equations of
motion are
$$
\nabla_t\partial_t\phi^i- h^{ij} \omega_{jk} \partial_t\phi^k=0\ ,
\eqno (2.2)
$$
where
$$
\omega=db\ ,
\eqno (2.3)
$$
$\nabla_i$ is the Levi-Civita covariant derivative of the metric $h$ and
$\nabla_t\equiv \partial_t\phi^i \nabla_i$.
The equations of motion are globally defined on $\cM$ because
$\omega\equiv db$ is a globally defined closed two-form, {\sl i.e.}\
$\omega_1=\omega_2$ on the
intersection of any two open sets $U_1$ and $U_2$ of $\cM$.

Let $G$ be a group and $f: G\times \cM\rightarrow \cM$
be a group action of $G$ on
 $\cM$; $f_{gg'}=f_g f_{g'}$ and $f_{e}=Id_{\cM}$ where
$g,g',e\in G$ ($e$ is the identity element of $G$).
Sufficient conditions for the
invariance of the equations of motion (2.2) under the group action $f$ are
$$
(f^*_g h)_{ij}=h_{ij}, \qquad (f^*_g \omega)_{ij}=\omega_{ij},\qquad g\in G\ ,
\eqno (2.4)
$$
{\sl i.e.}\   the transformations $f_g$ are
isometries and leave the closed two-form $\omega$
invariant.

In the quantum theory,
it is required that $\omega$ be the curvature of a line bundle over
the manifold $\cM$ and the wave
functions be sections of this line bundle; this property of
$\omega$  is a quantisation condition for the coupling
$b$ of the action (2.1) and it
is called Dirac's quantisation condition.  If in addition
this theory has symmetries
generated by a group action as above, the conditions (2.4)
are not enough to guarantee that
these symmetries can be implemented with unitary
transformations on the Hilbert space of the theory. For this, {\sl additional}
conditions are necessary.
To describe the additional conditions,  let $[\omega]$ be the
cohomology class of the curvature
$\omega$ in
$H^2(\cM, \ZN)$ and $G$ be compact and connected.  We first pull-back
$[\omega]$ using the group action $f$
on the manifold $G\times \cM$ and then decompose the
pulled-back cohomology class $f^*[\omega]$  as
$$
f^*[\omega]=[\omega]+[\sigma_1]+[\sigma_2]\ ,
\eqno (2.5)
$$
where $[\sigma_1]\in H^1(G, H^1(\cM,\ZN))$
and $[\sigma_2]\in H^2(G,\ZN)$. (We have used the
K{\"u}nneth formula to perform this decomposition).
Apart from (2.4), the {\sl additional}
conditions to implement the classical symmetries with unitary
transformations on the Hilbert space of the above theory are
$$
[\sigma_1]=0, \qquad [\sigma_2]=0\ .
\eqno (2.6)
$$
After some computation, we can
show that a consequence of the first condition of (2.6) is
that the $\exp(2 i \pi I)$ of
the action $I$ (eqn. (2.1)) and the charges of this theory
associated with the above symmetries
are globally defined on the manifold $\cM$, and a
consequence of the  second condition of (2.6)
is that the Poisson bracket algebra of these
charges is isomorphic to the Lie algebra of $G$.
For the proof of all the above statements
as well as the study of the case
where $G$ is disconnected using the theory of universal
classifying spaces see refs. [9].

In section 4, I will examine the analogue of the
first condition of (2.6) for the
case of two-dimensional sigma models
with WZ term.  It is worth pointing out though that
a condition similar to the second  of
(2.6) appears in the case of two-dimensional sigma
models as well but this will be presented elsewhere [10].

\bigskip
{\bf 3. Two-dimensional Sigma Models with Torsion}
\medskip

The action of a two-dimensional
sigma model with WZ term $b$ and  target space a Riemannian
manifold ($\cM,h$)  is
$$
I=\int d^2x (h+b)_{ij} \partial_\pp\phi^i \partial_=\phi^j -
V(\phi)\ ,
\eqno (3.1)
$$
where the sigma model fields $\phi$ are maps from the two-dimensional
space-time
$\Sigma$ with light-cone
co-ordinates $\{x^{\pp} =t+x, x^==t-x\}$ into $\cM$ and $V$ is a
real function on $\cM$.  The two-form $b$ is {\sl locally} defined on
$\cM$ with patching conditions
$b_1=b_2+dm_{12}$ at the intersection $U_1\cap U_2$ of any two open sets $U_1$,
$U_2$ of $\cM$ where $m_{12}$ is an
one-form defined on $U_1\cap U_2$.  The WZ term in the
action is then locally defined on
$\cM$.    One can define a closed three-form $H$  on
$\cM$ as
$$
H={3\over 2}db\ .
\eqno (3.2)
$$
Observe that $H$ is globally defined on $\cM$ and
it is called torsion for reasons that
will become apparent below.  The field equations are
$$
\nabla^{(+)}_=\partial_\pp\phi^i+{1\over2} h^{ij} \partial_jV=0\ ,
\eqno (3.3)
$$
where the connections of the covariant derivatives $\nabla^{(\pm)}$ are
$$
\Gamma^{(\pm)}{}^i_{jk}=\{^i_{jk}\}\pm H^i_{jk}\ .
\eqno (3.4)
$$
The tensor $H$ is the torsion of the connection
$\Gamma^{(+)}$ and it is globally defined on $\cM$.

Let $G$ be a group and $f: G\times \cM\rightarrow \cM$
be a group action of $G$ on
the target manifold $\cM$ as in the previous section; $f_{gg'}=f_g f_{g'}$ and
$f_{e}=Id_{\cM}$ where
$g,g',e\in G$.  Sufficient
conditions for the invariance of the field equations (3.3)
under the group action $f$ are
$$
(f^*_g h)_{ij}=h_{ij},
\qquad (f^*_g H)_{ijk}=H_{ijk}, \qquad f^*_g V=V,\qquad g\in
G\ ,
\eqno (3.5)
$$
{\sl  i.e.} the group action leaves
invariant the metric $h$ (so the group action $f$
generates isometries of
the Riemannian manifold $(\cM,h)$), the closed three-form
$H$ and the scalar potential $V$.   Let in addition $G$ be a Lie group with
Lie algebra $\cL(G)$.  The infinitesimal form of the above conditions  is then
$$
(L_ah)_{ij}=0, \qquad (L_aH)_{ijk}=0, \qquad  L_a V=0\ ,
\eqno (3.6)
$$
where $\{L_a; a=1,\dots, \rm{dim} \cL(G)\}$
is the Lie derivative with respect to the
vector field $\{X_a; a=1,\dots, \rm{dim}\ \cL(G)\}$
generated by the group action $f$
on $\cM$.  The vector fields $X_a$ are Killing ($\nabla_{(i}X_{j)a}=0$).

In the path-integral quantisation
of this theory, one needs to know the conditions for the
action (3.1) to be invariant
(up to an integer) under the transformations generated by a
group action.  The invariance
of the terms in the action (3.1) involving the metric $h$ and
the scalar potential $V$ follows
directly from the conditions on $h,V$ given in eqn. (3.5)
for the invariance of the
field equations.  However for the invariance of the Wess-Zumino
action in (3.1), we need additional
conditions besides those of eqn. (3.5) due partly to the
fact that this term is not globally
defined on $\cM$.  There are two main approaches to
define globally the Wess-Zumino term.
The first is the homotopy approach due to Wess and
Zumino [5], and  Rohm and Witten [6],
and, the second is the {\v C}ech Cohomology approach
due to O. Alvarez [7].  In
the next section, I will use the former to examine the symmetries
of the Wess-Zumino term.
A study of the symmetries of the WZ action in the  {\v C}ech
Cohomology approach will be presented elsewhere [10].

\bigskip
{\bf 4. Symmetries and the Wess-Zumino Action}
\medskip

Let  $\Sigma$, the two-dimensional
space-time,  be a closed manifold, {\sl i.e.}\   compact and
without boundary, and $[\Sigma, \cM]$
be the homotopy classes of maps from $\Sigma$ into the
sigma model target manifold $\cM$.
To define the Wess-Zumino term in the homotopy
approach, we  choose a \lq background'
map $\phi_0$ from $\Sigma$ into $\cM$ such that
$\phi_0$ is homotopic to $\phi$, {\sl i.e.}\
there is a map $F$ ($F: [0,1]\times \Sigma
\rightarrow \cM$) and
$$
 F(s,x)=\cases{\phi_0(x) &\qquad s=0\ ,\cr \phi(x) &\qquad s=1\ . }
\eqno (4.1)
$$
The homotopy $F$ interpolates
between $\phi_0$ and $\phi$.  In the following, we will use
the notation $\phi_0\eq_F\phi$ to
denote that the map  $\phi_0$ is homotopic to $\phi$ with
respect to $F$. The action of the Wess-Zumino term is defined [6] as
$$
S_{WZ}[\phi,\phi_0;F]=\int_{[0,1]\times \Sigma} F^*H\ .
\eqno (4.2)
$$
As indicated, the action of the Wess-Zumino term depends on the choice of the
homotopy $F$ that interpolates between  $\phi$ and $\phi_0$.  To determine the
dependence of $S_{WZ}$ on $F$, we take two different homotopies $F_1$ and $F_2$
that interpolate between $\phi$ and $\phi_0$ and compute the difference
$$
\Delta S_{WZ}=S_{WZ}[\phi,\phi_0;F_1]-S_{WZ}[\phi,\phi_0;F_2]\ .
\eqno (4.3)
$$
This difference can be rewritten as
$$
\Delta S_{WZ}=S_{WZ}[\phi_0,\phi_0;F_3]=\int_{S^1\times \Sigma} F_3^*H\ ,
\eqno (4.4)
$$
where
$$
F_3(s_3,x)=\cases{F_1(2s_3,x) \qquad \qquad  {0\leq s_3\leq {1\over 2}}
\cr
F_2(-2s_3+2,x) \qquad  {1\over 2}\leq s_3\leq 1\ .}
\eqno (4.5)
$$
Note that $F_3(0,x)=F_3(1,x)=\phi_0(x)$.
The difference
$\Delta S_{WZ}$ of (4.4) is the integral of a closed three-form over
a compact three-manifold without boundary and in general its value is a real
number.  However if $[H]\in H^3(M,\ZN)$, the difference is an integer and
thus the functional
$$
A[\phi,\phi_0]=e^{2i\pi S_{WZ}[\phi,\phi_0;F]}
\eqno (4.6)
$$
becomes independent of the choice of homotopy $F$. This property of $A$ is
{\sl
sufficient} for the
consistency of the path-integral quantisation of this theory.  In the
following, we will take
$[H]\in H^3(M,\ZN)$ and so the WZ action will
be independent from the choice of homotopy $F$
${\rm mod}\ 1$.

The invariance of the field
equations of a sigma model with a Wess-Zumino term under
the group action  $f$ of a
(connected) group $G$ on $\cM$  does not necessarily imply the
invariance of the
action $S_{WZ}[\phi,\phi_0;F]$.  The transformation $\phi^g\equiv
f_g(\phi)$ of the field $\phi$
induces the transformation $S_{WZ}[\phi^g,\phi_0;F_4]$
on the Wess-Zumino action which can be rewritten as
$$\eqalign{
S_{WZ}[\phi^g,&\phi_0;F_4]=S_{WZ}[\phi^g,\phi_0;F_3] +{\rm integer}
\cr
=&S_{WZ}[\phi^g,\phi^g_0;F_2]+S_{WZ}[\phi_0^g,\phi_0;F_1] +{\rm integer}\ ,}
\eqno (4.7)
$$
where
$$
F_3(s,x)=\cases{F_1(2s,x) \qquad \qquad 0\leq s \leq{1\over 2}\ ,
\cr
F_2(2s-1,x)\ \qquad {1\over 2}\leq s \leq 1\ .}
\eqno (4.8)
$$
The first equality in (4.7) follows
from the observation that $\phi_0\eq_{F_3}\phi^g$ and
the property of the WZ action to be
independent of the choice of homotopy $F$ up to an
integer, and the second from the
definition of the WZ term. Using again the property of the
WZ action to be independent of the choice of homotopy
${\rm mod}\ 1$, we can choose $F_2=F^g$ and
then use the property of $H$ to be invariant
under the group action $f$
(this comes from the invariance of the field equations eqn.
(3.5)) to reexpress  (4.7) as
$$
S_{WZ}[\phi^g,\phi_0;F_4]=S_{WZ}[\phi,\phi_0;F]+S_{WZ}[\phi_0^g,\phi_0;F_1] +
\rm{integer'}\ .
\eqno (4.9)
$$
{}From (4.9) it is clear that in {\sl addition} to the conditions (3.5) for the
invariance of the field equations, the vanishing  of
$$
c[\phi_0,g]= S_{WZ}[\phi_0^g,\phi_0; F_1]\quad {\rm mod}\  1\ .
\eqno (4.10)
$$
is a necessary condition [8] in order for
the group action $f$ of $G$ to be a symmetry of the
action
$S_{WZ}[\phi,\phi_0;F]$.
Note that $c[\phi_0,g]$ is independent of the choice of homotopy
$F_1$.
 I will refer to $c[\phi_0,g]$ as anomaly.

The anomaly $c[\phi_0,g]$ has some
novel properties.  In particular, we can show that
$c[\phi_0,g]$ depends on the
homotopy class $[\phi_0]$ of $\phi_0$ rather than $\phi_0$
itself.  To prove this, let $\phi_1\eq_{F_1}\phi_0$. We
write
$$\eqalign{
S_{WZ}[\phi_0^g,&\phi_0;F]=S_{WZ}[\phi_0^g,\phi_0;F_5]+ {\rm integer}
\cr
=&S_{WZ}[\phi_0^g,\phi_1^g;F_2]+S_{WZ}[\phi_1^g,\phi_1;F_3]+
                S_{WZ}[\phi_1,\phi_0;F_4]+{\rm integer}\ ,}
\eqno (4.11)
$$
where
$$
F_5(s_5,x)=\cases{F_4(3 s_5, x)\qquad \qquad 0\leq s_5\leq {1\over 3}\ ,
\cr
F_3(3 s_5-1, x)\ \qquad {1\over 3}\leq s_5\leq {2\over 3}\ ,
\cr
F_2(3 s_5-2, x)\ \qquad {2\over 3}\leq s_5\leq 1\ .}
\eqno (4.12)
$$
Observe from the expression for $F_5$ that $\phi_0\eq_{F_5} \phi_0^g$.
Using the property of the
Wess-Zumino action to be independent of the choice of homotopy
${\rm mod}\ 1$ and the invariance of $H$, we write
$S_{WZ}[\phi_0^g,\phi^g;F_2]=S_{WZ}[\phi_0,\phi;F_1]+{\rm {integer}}$.
The equation (4.11)
then becomes
$$\eqalign{
S_{WZ}[\phi_0^g,\phi_0;F]=&S_{WZ}[\phi_0,\phi_1;F_1]+S_{WZ}[\phi_1^g,\phi_1;F_3]
\cr &
+S_{WZ}[\phi_1,\phi_0;F_4]+ {\rm integer'}
\cr
=&S_{WZ}[\phi_1^g,\phi_1;F_1]+{\rm integer''}\ .}
\eqno (4.13)
$$
So if $\phi_0$ is homotopic
to $\phi_1$, we have shown that $c[\phi_1,g]=c[\phi_0,g]$ and
thus the anomaly  $c$ is a map from $[\Sigma, \cM]\times G$ into $\RN/\ZN$.

We can also show that
$$
c\big{[}[\phi], g_1g_2\big{]}
= (c\big{[}[\phi], g_1\big{]}+c\big{[}[\phi],g_2\big{]}){\rm
mod\  1},
\qquad g_1, g_2\in G\ .
\eqno (4.14)
$$
This immediately follows from
$$
\eqalign{S_{WZ}[\phi_0^{g_1g_2}&,\phi_0;F]=
S_{WZ}[\phi_0^{g_1g_2},\phi_0;F_3]+{\rm
integer}
\cr=&S_{WZ}[(\phi_0^{g_1})^{g_2},\phi^{g_2}_0;F_2]
+S_{WZ}[\phi_0^{g_2},\phi_0;F_1] +{\rm
integer}
\cr
=&S_{WZ}[\phi_0^{g_1},\phi_0;F_4]
+S_{WZ}[\phi_0^{g_2},\phi_0;F_1]+{\rm integer'}\ ,}
\eqno (4.15)
$$
where $F_3$ is related to $F_1$ and $F_2$ as in (4.8) and $F_2=F^{g_2}_4$.
In addition,
$c\big{[}[\phi], e\big{]}=0$, so the anomaly is represented by a group
homomorphism from $G$
into $\RN/\ZN$ for every homotopy class of maps from $\Sigma$ into
$\cM$.

An important consequence of the above properties of the anomaly $c$ is that $c$
vanishes whenever it is evaluated
on the trivial topological sector of the theory, {\sl
i.e.} on the trivial class of $[\Sigma,\cM]$.  To prove this, we observe that
the trivial topological sector
can be represented by the constant maps from $\Sigma$ into
$\cM$.  Since the anomaly
$c$ depends on the homotopy classes
of the maps rather than the maps themselves, we can use
any representative of the trivial homotopy class, hence any constant map from
$\Sigma$ into $\cM$, to do the
computation.  Let $\phi_0$ be such a map,  the expression
(4.10) for the anomaly involves
the pull-back $F^*H$ of the three-form $H$ with respect to
the homotopy $F$ that interpolates between
$\phi_0^g$ and $\phi_0$.
Since $\phi_0^g$ and $\phi_0$ are constants maps from $\Sigma$
into $\cM$, they can be thought of as points in $\cM$ and so we can take
$F$ to be any path in $\cM$ that
interpolates between them.  In which case, $F$ is a map of
one variable and therefore
the pull-back form $F^*H$ of the three-form  $H$ vanishes
($F^*H=0$). In fact using the same arguments, we can show that the
anomaly $c$ vanishes for all the
topological sectors $[\phi]$ of the theory that admit a
homotopy $F$ which interpolates
between $\phi$ and $\phi^g$ with differential map
$dF: T([0,1]\times \Sigma)\rightarrow T(\cM)$
that has rank less than three. Thus  only the
{\sl non-trivial} topological sectors
can break the symmetries of the field equations of a
sigma model with a WZ term.

The anomaly $c$ vanishes for all sigma
models with torsion for which $[\Sigma, \cM]$ has
only one element, the trivial class.
For  example, this is the case for the sigma models
with $\Sigma=S^2$ and $\cM=SU(N)$ because $\pi_2(SU(N))=0$.

It is well known that the homotopy
classes of maps $[\Sigma, \cM]$ under certain conditions
can be given a group structure.
So it is natural to ask whether or not the anomaly $c$ is a
group homomorphism from
$[\Sigma, \cM]\times G$ into $\RN/\ZN$.  We have already shown
above that $c\big{[}[\phi],g\big{]}=0$
if $[\phi]$ is the trivial class of $[\Sigma, \cM]$.
So it remains to show whether $c\big{[}[\phi_1\cdot\phi_2
],g\big{]}=(c\big{[}[\phi_1],g\big{]}+c\big{[}[\phi_2],g\big{]})\  \rm {mod}
\ 1$.  Examination of the particle
model in section 2 indicates that indeed the anomaly $c$
is a group homomorphism but this will not be pursued further here.

Next we will calculate the form of the
anomaly for infinitesimal transformations.  For this,
let us consider an one-parameter subgroup of $G$ generated by the vector
$v\in \cL(G)$ . The elements of
this subgroup can be written as $g(r)={\rm{exp}}( r v)$ where
$r$ is a real number and
$g$ can be thought as a map from the real line $\RN$ into $G$.
Using this subgroup,
we define another map that we will call again  $g$ from $[0,1]\times
(-\epsilon,\epsilon)$
into $G$ by setting $r=sz$, {\sl i.e.}\   $g(s,z)={\rm exp}( sz v)$,
where $\epsilon$ is a
positive real number. We observe that $g(s,z)$ has the following
properties:
$$
g(s,z)=\cases{e &\quad \rm{if\ \ \ either}\quad s=0\ \ \rm{ or}\quad  z=0\cr
g(z)={\rm{exp}}z v &\quad  s=1\ .}
\eqno (4.16)
$$
The \lq infinitesimal' anomaly is given by
$$
\Delta[\phi,v]:={d\over dz}S_{WZ}[\phi^{g(z)},\phi;F]|_{z=0}\ ,
\eqno (4.17)
$$
where
$$
F(s,x):=\phi^{g(s,z)}(x)\ .
\eqno (4.18)
$$
Because of (4.16), the homotopy
$F$ interpolates between $\phi$ and $\phi^{g(z)}$.
Using the definition (4.17) of
the infinitesimal anomaly and the Wess-Zumino action, we can
show after some further computation that
$$
\Delta[\phi,v]=\int_{\Sigma} v^a \phi^*(\eta_a)\ ,
\eqno (4.19)
$$
where
$$
\eta_a\equiv i_{a} H\ ,
\eqno (4.20)
$$
and $i_a$ is the inner derivation with respect to the Killing
vector field $X_a$.
Using  $dH=0$ from (3.2),
$L_aH=0$ from (3.6) and  $L_a=di_a+i_ad$, we can show that
$\eta_a\equiv i_a H$ is  a closed two-form
$$
d\eta_a=0\ .
\eqno (4.21)
$$
Therefore the infinitesimal anomaly $\Delta[\phi,v]$ is the integral of the
pull-back with the map $\phi$ of a
closed two-form of $\cM$ with domain the space-time
$\Sigma$ and so it depends on
the homotopy class $[\phi]$ of $\phi$, {\sl i.e.}\
$\Delta=\Delta\big{[}[\phi],v\big{]}$.

The infinitesimal anomaly
$\Delta\big{[}[\phi],v\big{]}$ vanishes whenever $[\phi]\in
[\Sigma,\cM]$ is the trivial class (in agreement with the behaviour of
$c\big{[}[\phi],g\big{]}$
discussed above).  The anomaly $\Delta\big{[}[\phi],v\big{]}$ also
vanishes provided that the two-forms
$\{\eta_a
\}$ are exact.
Note though that $\Delta\big{[}[\phi],v\big{]}$ does {\sl not} vanish if
 $b$ is merely globally defined on $\cM$.

The Noether charges associated
with the symmetries of the action (3.1) generated by
the group action $f$ are
$$
Q_a=\int\  dx\  (X_{ai} \partial_t\phi^i+u_{ai} \partial_x\phi^i)\ ,
\eqno (4.22)
$$
where the one-forms $\{u_a\}$ are
locally defined and they are related to the closed
two-forms
$\{\eta_a\}$ as follows:
$$
\eta_a=du_a\ .
\eqno (4.23)
$$
If the two-forms $\{\eta_a\}$ are exact, then $\{u_a\}$ are globally
defined and consequently the
Noether charges $\{Q_a\}$ are globally defined as well.
However even if $\{u_a\}$ are globally
defined, it is not expected that the
Poisson bracket algebra of these charges  to be
necessarily
isomorphic to $\cL(G)$.  Recall that the Poisson bracket algebra of the charges
of the particle model in section 2 has similar behaviour.

The methods developed above to examine the symmetries  of the WZ
action for two-dimensional
sigma models can be extended to the case of n-dimensional ones
[8].

\bigskip
{\bf 5. Concluding Remarks}
\medskip

Symmetries generated by
vector fields on the sigma model manifold arise naturally in
the study of (p,q)-supersymmetric
two-dimensional massive sigma models because the charges of
such symmetries appear as
central charges in the Poisson bracket algebra of supersymmetry
charges of these models.  The
above vector fields also enter in the expression for the scalar
potential.

The simplest massive supersymmetric
sigma model with a central charge is the one with
(1,1)-supersymmetry.
Let $X$ be a Killing vector field that leaves invariant both the
torsion $H$ of eqn. (3.2)
and $u$ of eqn. (4.23). It can be shown that the most general
action of a massive sigma model with (1,1) supersymmetry [3] is
$$
\eqalign{
I =\int\! d^2 x\big\{ &\partial_\pp\phi^i\partial_=\phi^j (h_{ij}+b_{ij})
+ ih_{ij}\lambda_+^i\nabla_=^{(+)}\lambda_+^j -
ih_{ij}\psi_-^i\nabla_\pp^{(-)}\psi_-^j\cr
& -{1\over2}\psi_-^k\psi_-^l\lambda_+^i\lambda_+^j R^{(-)}_{ijkl}
 +m\nabla^{(-)}_i (u -X)_j \lambda_+^i \psi_-^j  -V(\phi) \big\}\ ,}
\eqno (5.1)
$$
where  $\lambda_+$ and $\psi_-$ are real chiral fermions and
$$
V(\phi)={m^2\over 4} h^{ij} (u-X)_i (u-X)_j\ ,
\eqno (5.2)
$$
is the scalar potential.
The rest of the notation follows from sections 3 and 4.  Observe
that the requirement
for the action (5.1) to be (1,1)-supersymmetric imposes strong
restrictions on the form of the scalar potential
$V(\phi)$.

To define globally the
field equations derived from this action on the sigma model target
space $\cM$,  we have to
assume that $u$ is globally defined one-form on $\cM$.  As we have
established in the previous
section, this is a sufficient condition for the WZ  action to be
invariant under the infinitesimal transformations generated by $X$.

Computation of the Poisson
bracket algebra of charges of the above model reveals that
$$
\{S_+,S_+\}= E+P, \qquad \{S_-,S_-\}=E-P, \qquad \{S_+,S_-\}=Q_X\ .
\eqno (5.3)
$$
where $S_+, S_-$ are the
supersymmetry charges, $E$ is the energy, $P$ is the momentum and
$Q_X$ is the Noether charge
associated with the symmetries generated by $X$ (eqn.
(4.22)). To derive (5.3),
we have assumed $X^i u_i=0$.  Observe that $Q_X$ appears in the
Poisson bracket of left- with
right- supersymmetry charges.  The form of the charges as well
as details of this computation can be found in refs. [4].

In conclusion, non-trivial
topological sectors can break quantum mechanically the symmetries
of the field equations of a sigma model with torsion.  This is partly due to
the
WZ action which may not be globally defined and/or may be invariant
up to surface terms
that cannot be integrated away.  An associated anomaly is computed and
it is found to be
a WZ-like action that depends on the homotopy class of the topological
sector of the theory and
the group action on the sigma model manifold that generates the
symmetries of the field
equations.  Sufficient conditions for the vanishing of this anomaly
are also given.
In particular, it is shown that the anomaly vanishes whenever the theory
has only one topological
sector, {\sl i.e.}\  the trivial one.  Finally it is worth pointing
out  that, as in the case
for the one-dimensional sigma model, other conditions are
necessary in addition to
the vanishing of the above anomaly in order for the symmetries of
the field equations of
n-dimensional ($n\geq 2$) sigma models with torsion to be implemented
quantum mechanically.

\bigskip

\noindent{\bf Acknowledgements:} Discussions with C.M. Hull,
H. Nicolai, P.K. Townsend
and A. van de Ven are gratefully
acknowledged. I was funded by a grant from the European
Union.

\hfill\eject
\centerline {\bf References}
\medskip

\item {[1]}
T.H. Buscher, Phys. Lett. {\bf 159B} (1985) 127; Phys. Lett. {\bf 194B} (1987)
51.
M. Ro{\v c}ek and E. Verlinde, Nucl. Phys. {\bf B373} (1992) 630.
A. Giveon and M. Ro{\v c}ek, Nucl. Phys. {\bf B380} (1992) 128.
E. Alvarez, L. Alvarez-Gaum{\' e},
and Y. Lozano, CERN-TH-7204/94, hep-th/9403155.

\item {[2]}
 L. Alvarez-Gaum{\' e} and D.Z. Freedman, Commun. Math. Phys. {\bf 91} (1983)
87.

\item {[3]}
C.M. Hull, G. Papadopoulos and P.K. Townsend, Phys. Lett. {\bf 316B} (1993)
291.

\item {[4]}
G. Papadopoulos and P.K. Townsend, Class. Quantum Grav., Vol. 2,
{\bf 3} (1994) 515; \lq \lq Massive
(p,q)-supersymmetric sigma models revisited"
   R/94/9, DESY 94-092, hep-th/9406015.

\item {[5]}
J. Wess and B. Zumino, Phys. Lett {\bf 37B} (1971) 96.

\item {[6]}
R. Rohm and E. Witten, Ann. Phys. {\bf 170} (1986) 454.

\item {[7]}
O. Alvarez, Commun. Math. Phys. {\bf 100} (1985) 279.

\item {[8]}
G. Papadopoulos, Class. Quantum Grav. {\bf 7} (1990) L41.

\item {[9]}
G. Papadopoulos, Commun. Math.
Phys. {\bf 144} (1992) 491; Phys. Lett. {\bf 248B } (1990)
113.

\item {[10]}
C.M. Hull and G. Papadopoulos, in preparation.
\bye

\end